\documentstyle[prl,aps,epsf,array]{revtex}
\begin{document}
\draft

\title{Kondo Resonance in a Mesoscopic Ring Coupled to a
Quantum Dot: \\
Exact Results for the Aharonov-Bohm-Casher Effects}

\author{H.-P.\ Eckle,$^{1,2}$ H.\ Johannesson,$^3$ and C.\ A.\ Stafford$^4$}
\address{$\mbox{}^1$School of Physics, The University of New South Wales,
Sydney, 2052, Australia}
\address{$\mbox{}^2$Department of Physics, University of Jyv\"askyl\"a,
FIN--40351 Jyv\"askyl\"a, Finland}
\address{$\mbox{}^3$Institute of Theoretical Physics,
Chalmers University of Technology and G\"oteborg University,
SE 412 96 G\"oteborg, Sweden}
\address{$\mbox{}^4$Department of Physics, University of Arizona, 
1118 E.\ 4th Street, Tucson, AZ 85721}

\twocolumn[\hsize\textwidth\columnwidth\hsize\csname@twocolumnfalse\endcsname

\maketitle

\begin{abstract}

We study the persistent currents
induced by both the Aharonov-Bohm and Aharonov-Casher effects 
in a one--dimensional mesoscopic ring coupled to
a side--branch quantum dot at Kondo resonance.
For privileged values of the Aharonov-Bohm-Casher
fluxes, the problem can be mapped
onto an integrable model, exactly solvable by a Bethe ansatz.
In the case of a pure magnetic Aharonov-Bohm flux, we find
that the presence of the quantum dot has no effect on the persistent
current.
In contrast, the Kondo resonance
interferes with the spin-dependent
Aharonov-Casher effect to induce a current which,
in the strong-coupling limit, is independent of
the number of electrons in the ring.

\end{abstract}

\vskip0pc]

The Kondo effect---where the interaction
between a local spin and free electrons produces a strongly-correlated 
state below a characteristic temperature $T_K$---has
become one of the paradigms in the study of correlated electron
behavior \cite{Hewson}.
In a recent experimental breakthrough
\cite{Goldhaber}, a tunable Kondo effect was observed in
ultra small semiconductor quantum dots connected capacitively to a
gate and via tunnel junctions to electrodes.
By sweeping the gate
voltage, the dot's highest spin-degenerate level $\epsilon_{d}$ 
can be tuned relative to the chemical potential $\mu$ of the
leads.
This level is occupied by a single electron when
$\epsilon_{d}\le\mu-\Gamma_{d}$, with $\Gamma_{d}=2\pi
\sum_k |V_k|^2\delta(\epsilon_{d} - \epsilon_k)$ the one-particle
resonance width of the dot, and $V_k$ the 
tunneling matrix elements
through the junction barriers.
Below a temperature $T_K \sim
\mbox{exp}(-\pi|\mu-\epsilon_{d}|/\Gamma_{d})$ the resulting free
spin on the dot forms a singlet with the electron spins in the leads
via virtual co-tunneling processes.
A fingerprint of this strongly correlated 
state is the dramatic enhancement of the local spectral
density at the Fermi level.
As predicted theoretically\cite{Glazman,Meir}
and as seen in the experiments\cite{Goldhaber}, 
this makes the dot transparent to electron
transport when $T \ll T_K$.

An interesting problem is how the {\em persistent current} (PC) of a
multiply connected system coupled to a quantum dot is affected by
a Kondo resonance. 
A PC is the equilibrium response \cite{by,BIL,ZK} 
to a magnetic Aharonov-Bohm (AB) flux \cite{AB} and/or an
Aharonov-Casher ``flux'' \cite{AC} of a charged wire piercing the
system.  In contrast to ordinary (nonequilibrium) currents, as
measured in the experiments referred to above, a PC
requires for its existence that an electron maintains its 
{\em phase coherence} while circling the ring, and should thus be
sensitive to scattering in the quantum dot \cite{Yacoby}.
The PC of a one-dimensional (1D) ring coupled to a quantum dot
was previously investigated by B\"uttiker and one of the authors 
\cite{mbcs}, considering two different topologies: one in which 
the electron had to tunnel through the quantum dot in order to encircle
the flux (embedded dot), 
and one in which  an intact ring was coupled via
a tunnel barrier to an adjacent quantum dot
(sidebranch dot).  However, in Ref.\ \onlinecite{mbcs}, the energy level
spacing $\delta$ in the mesoscopic ring was taken to be much greater than 
$T_K$, so that although a singlet state was formed between
the electron on the dot and those within the ring, 
this state had more in common with a (two-body)
chemical bond than with a true (many-body) Kondo resonance: 
the finite-size effects implied by $\delta \gg T_K$ here suppress the Kondo effect.  

Recently, the problem of a ring with an embedded quantum dot has been
revisited by a number of authors \cite{Ferrari,Kang,Deo}, and the Kondo 
scaling with $\delta$ has been investigated \cite{Kang}.  
Based on a cluster calculation
in the grand canonical ensemble, Ferrari et al.\ \cite{Ferrari} claim that
the PC is strongly enhanced by the Kondo effect, scaling as
$L^{-1/2}$ when $\delta > T_K$ (this includes 
the regime studied in Ref.\ \onlinecite{mbcs}, where no appreciable enhancement
was found).
In contrast, using a variational approach, Kang and Shin \cite{Kang}
find that the Kondo-assisted PC is comparable to that of a perfect ring
for an even number of electrons, and strongly suppressed for an odd number of
electrons.
Other related results have also been obtained \cite{Deo,ZB}.
Clearly, the situation calls for a more 
definitive analysis to clarify the underlying physics of this class of problems.

In this Letter, we investigate the PC in a 1D mesoscopic ring coupled via tunneling to
a sidebranch quantum dot in the Kondo regime, with $\delta \leq T_K$ 
(opposite to the limit $\delta \gg T_K$ considered in \onlinecite{mbcs}).
We exploit the fact that the
PC to leading order in 1/L effectively involves only processes close to the 
Fermi level \cite{Cheung,Loss}:
In a 1D ring, the system responds to the AB and/or AC
fluxes by virtually creating and annihilating states at the left and
right Fermi points, and this leads to a change in the energy on a
mesoscopic scale, producing a PC. Since the coupling to the dot via
the Kondo exchange influences only states at the Fermi level, this
allows us to construct an {\em effective} model valid in the neighborhood 
of the Fermi points.
The model turns out to be exactly solvable for a
discrete set of values of the AB and/or AC fluxes, for which
we  can 
calculate the PC {\em exactly}.  Via adiabatic continuation, we conclude 
that the AB PC is {\em insensitive} to the
presence of the quantum dot for {\em any} weak 
AB flux,  
thus extending the result found in \cite{mbcs} into the Kondo regime. 
In contrast, with an AC flux we find that the Kondo resonance produces an imbalance between right- and left moving charge carriers, leading to a net AC PC. 

A quantum dot connected to a 1D electron
reservoir (in our case, a mesoscopic ring) is conventionally 
described by the Anderson impurity 
model \cite{Glazman,Meir}: 
\[
H = \sum_{\alpha} \epsilon_{d}d^{\dagger}_{\alpha}d_{\alpha} +
U n_\uparrow n_\downarrow +
\sum_{k, \alpha} \epsilon_k c^{\dagger}_{k\alpha}c_{k\alpha}
+ (V_k c^{\dagger}_{k\alpha}d_{\alpha}
+ \mbox{H.c.}), 
\]
where 
$d^{\dagger}_{\alpha}$ creates an electron on the dot with spin $\alpha$,
$c^\dagger_{k\alpha}$ creates an electron of energy $\epsilon_k$ in the ring,
and $n_\alpha =d^\dagger_\alpha d_\alpha$.
Here we consider only a single spin-degenerate level $\epsilon_{d}$ within
the quantum dot. We shall take the charging energy $U \rightarrow \infty$, as is appropriate for quantum dots where $U \sim 1$ meV and $\Gamma_d \sim 1-10 
\mu$eV \cite{footnote}. 

When $\epsilon_{d}\ll\mu-\Gamma_{d}$, the dot is singly occupied,
i.e., $\sum_\alpha \langle d^{\dagger}_{\alpha}d_{\alpha} \rangle = 1$,
and the Anderson model can be
mapped onto a Kondo Hamiltonian via a Schrieffer-Wolff
transformation \cite{Hewson}. To have a faithful representation we require that
$\delta \le T_K$, which ensures a finite DOS at the Fermi level and hence
a fully developed Kondo resonance \cite{Thimm}. Passing to a real-space
continuum description, we obtain
\begin{eqnarray}
H & = & - \frac{\hbar^2}{2m_e} 
\sum_\alpha \int_0^L dx \psi_{\alpha}^{\dagger}(x)
\partial^2_x \psi_{\alpha}(x) 
\nonumber \\
& & \mbox{} + \lambda \sum_{\alpha,\beta}
\psi_{\alpha}^{\dagger}(0)
\mbox{\boldmath $\sigma$}_{\alpha \beta} \psi_{\beta}(0) \cdot 
\mbox{\boldmath $S$},
\end{eqnarray}
where the antiferromagnetic spin exchange takes the value
$\lambda \sim |V_{k_F}|^2/|\epsilon_{d}|$ 
(with the chemical
potential $\mu$ taken as the zero of energy), $L$ is the circumference of the
ring, $\mbox{\boldmath $S$}$ is the spin-1/2 operator of the electron in
the dot (located at $x=0$),
and $\psi_{\alpha}(x)$ is an electron field with spin index $\alpha = \pm 1$.
Next, we thread the ring with a magnetic flux $\Phi$, producing an
Aharonov-Bohm effect \cite{AB}.
In addition, to probe the {\em
spin-dependent} equilibrium response of the system, we pierce the ring with a
charged wire of charge $\tau$ per unit length. This will cause an
Aharonov-Casher effect \cite{AC} driven by the relativistic spin-orbit
interaction due to the electric field from the wire.
The quantum phases from the combined
Aharonov-Bohm-Casher (ABC) effects can be encoded, via a gauge
transformation \cite{by}, in the twisted boundary conditions $\psi_{\alpha}(L)
= e^{i\phi_{\alpha}}\psi_{\alpha}(0)$, where
\begin{equation}
\phi_{\alpha} = 2\pi(\frac{\Phi}{\Phi_0} + \alpha
\frac{\tau}{\tau_0}), \ \ \ \alpha = \pm 1.   \label{ABC}
\end{equation}
Here
$\Phi_0 = hc/e$ is the elementary magnetic flux quantum and
$\tau_0 = hc/\mu_e$ its electromagnetic dual, $\mu_e$
being the electron magnetic moment.

Since the essential physics is confined to a small
region around the left and right Fermi points, 
we can linearize the spectrum around $\pm k_{\mathrm{F}}$ \cite{Loss,Thimm},
and introduce left ($l$) and right ($r$) moving chiral fields:
\begin{equation}
\psi_{\alpha}(x) \sim \mbox{e}^{-ik_{\mathrm{F}} x}\psi_{l,\alpha}(x) +
                   \mbox{e}^{ik_{\mathrm{F}} x}\psi_{r,\alpha}(x).
\label{eq:lrm}
\end{equation}
Furthermore we introduce a basis of definite parity 
fields ({\em Weyl basis}):
\begin{equation}
\psi_{\mathrm{e/o},\alpha}(x) = 
\frac{1}{\sqrt{2}}\left(\psi_{r,\alpha}(\pm x) \pm 
                                        \psi_{l,\alpha}(\mp x)\right),
\label{eq:Weyl1}
\end{equation}
with $\psi_{\mathrm{e}}$ an even--parity, right--moving electron field, and
$\psi_{\mathrm{o}}$ an odd--parity, left--moving field. 
In the Weyl basis, the linearized Hamiltonian takes the form
$H = H_0^{\mathrm{odd}} + H_0^{\mathrm{even}}
                       + H_{\mathrm{imp}}^{\mathrm{even}}$, 
where
\begin{equation}
H_0^{\mathrm{even/odd}}   = 
\mp \frac{v_{\mathrm{F}}}{2\pi}  \sum_\alpha
\int_0^L dx 
                                    \psi_{\mathrm{e/o},\alpha}^\dagger(x)
                                i\partial_x \psi_{\mathrm{e/o},\alpha}(x)
\end{equation}
describe free Dirac electrons with velocity $v_{\mathrm{F}}$, and
the impurity contribution is diagonal in the channel index:
\begin{equation}
H_{\mathrm{imp}}^{\mathrm{even}} = \lambda \sum_{\alpha,\beta}
                                       \psi^\dagger_{\mathrm{e},\alpha}(0)
                                       \mbox{\boldmath $\sigma$}_{\alpha \beta}
                                       \psi_{\mathrm{e},\beta}(0) \cdot
                                       \mbox{\boldmath $S$}. \label{kondo}
\end{equation}
We recognize $H_K^{\mathrm{even}} \equiv 
              H_0^{\mathrm{even}} + H_{\mathrm{imp}}^{\mathrm{even}}$ 
as the chiral Hamiltonian of the spin-$1/2$ Kondo model \cite{Hewson}. 

While the even and odd parity channels are decoupled {\em in the Hamiltonian},
they become connected again by the twisted boundary conditions:
\[
\left(\begin{array}{c} \psi_{{\rm e},\alpha}(L) \\ 
\\
\psi_{{\rm o},\alpha}(L) \end{array}\right) =
\left(\begin{array}{cc} \cos\phi_\alpha & i\sin\phi_\alpha  \\
\\
-i\sin\phi_\alpha  & \cos\phi_\alpha \end{array}\right)
\left(\begin{array}{c} \psi_{{\rm e},\alpha}(0) \\ 
\\
\psi_{{\rm o},\alpha}(0) \end{array}\right), 
\]
where in (\ref{eq:lrm}) we have taken $k_{\mathrm F}=(2\pi/L)n$, 
with $n$ an integer.    
At the special values $\phi_\alpha=f_\alpha\pi$, with
$f_\alpha$ an integer, however,
this matrix reduces to a multiple
of the unit matrix, and the even and odd parity states decouple from each
other entirely.  For the corresponding values of the ABC
fluxes, the model becomes integrable and $H_K^{\rm even}$ 
can be solved by a Bethe ansatz \cite{baKondo}.
Thus, our original problem has been mapped onto an exactly
solvable problem consisting of a left-moving odd-parity branch of free
Dirac electrons, together with a (decoupled) right-moving 
even-parity branch, described by an integrable 1D Kondo model.

Having constructed an effective model valid near the Fermi level, we need to
properly define the (charge) PC of the system
using only Fermi level properties. This can be done as follows:
We introduce the {\em excess numbers}
$\Delta N_{r/l}$ of particles (holes) on the right/left dispersion branches
due to the Aharonov-Bohm-Casher fluxes:
\[
\Delta N_{r/l}(\phi_{\alpha}) = \frac{L}{2\pi}
                    [|k_{r/l,{\mathrm F}}(\phi_{\alpha})|-
{\mathrm min}(|k_{l,{\mathrm F}}(0)|, k_{r,{\mathrm F}}(0))],
\]
with $k_{r/l,{\mathrm F}}$ flux-dependent
momenta associated with the highest occupied level on the
respective branch. (For free electrons,
$k_{r/l,{\mathrm F}}=
[\pm 2\pi n_{r/l,{\rm max}}+\phi_{\alpha}]/L$).
This correctly accounts for the {\em average} number
of surplus current-carrying particles ($\Delta N_{r/l} > 0$) and holes 
($\Delta N_{r/l} < 0$) due to the fluxes when the
total particle number $N$ is fixed and hence the chemical potential
is free to vary between the highest occupied and lowest unoccupied levels.
Since the charge velocity
$v_{\rm F}$ is unrenormalized by interactions in this
model\cite{baKondo}, 
each current-carrying state contributes with $\pm ev_{\rm F}/L$ to 
$I(\Phi,\tau)$ (with the sign determined by the branch $r/l$
and the sign of $\phi_{\alpha}$), and we thus obtain
\begin{equation}
I(\Phi,\tau) = 
-\frac{ev_{\rm F}}{L}\sum_{\alpha =\pm 1}(\Delta N_{r}(\phi_{\alpha})
 - \Delta N_{l}(\phi_{\alpha})).
\label{PersistentCurrent}
\end{equation}
When the total number $N$ of electrons in the ring is even (which, for
simplicity,
is the case we shall focus on below), we still have to distinguish two cases:
When $N = 4n+2$, with $n$ an integer, $k_{r,F}(0) = |k_{l,F}(0)|$, while for 
$N = 4n$ the lifting of the
groundstate degeneracy due to the flux in this case implies that
$k_{r,F}(0_{\pm}) = |k_{l,F}(0_{\pm})| \mp 2\pi/L$ in the limit $\phi_{\alpha}
\rightarrow 0_{\pm}$. This leads to the well-known parity
effect \cite{by,Loss,ZK} 
(here for an even number of spinful electrons), as is easily verified from
(\ref{PersistentCurrent}).

Given (\ref{PersistentCurrent}), 
the problem is now reduced to calculating how the
{\em excess numbers} 
depend on the ABC
fluxes in the presence of a Kondo dot. To do this, we apply 
the techniques of the {\em Bethe ansatz} for finite systems,
developed previously for the 1D Hubbard
model~\cite{we}.
As we have already noted, our model is integrable for 
$\phi_\alpha=f_\alpha\pi$, with 
$f_\alpha$ an integer.
The nested Bethe ansatz equations (BAE) which diagonalize $H$ in this case 
are 
\[
Lk_{n_l}= -2\pi n_{l} + f_c\pi 
+\left(\frac{2M_{\rm o}}{N_{\rm o}} - 1\right)f_s\pi
+ \frac{2\pi}{N_{\rm o}}
\sum_{\delta=1}^{M_{\rm o}} J_\delta,
\]
\[
Lk_{n_r} = 2 \pi n_{r} + f_\downarrow\pi + \sum_{\gamma = 1}^{M_{\rm e}} 
                 \left[\theta(2\Lambda_\gamma - 2) - \pi \right],
\]
\[
N_{\rm e}
 \theta(2\Lambda_\gamma - 2) + \theta(2\Lambda_\gamma) 
= 2 \pi I_\gamma
+(f_\uparrow - f_\downarrow)\pi
 + \sum_{\delta = 1}^{M_{\rm e}} \theta(\Lambda_\gamma -\Lambda_\delta ),
\]
where $k_{n_l}$ are the pseudomomenta characterizing
the  $N_{\rm o}$ odd-parity {\em left} movers 
which decouple from the impurity, 
$M_{\rm o}$ of which have spin down,
and $k_{n_r}$ are pseudomomenta characterizing
the $N_{\rm e}$ even-parity {\em right} movers, 
with $M_{\rm e}$ counting the number of down spins in this sector. The 
quantum numbers $n_l, n_r,
I_{\gamma}$ and $J_{\delta}$ specifying the state take integer or half-odd integer
values
depending on the values of $M_{\rm e/o}$ and $N_{\rm e/o}$,
while $\{\Lambda_\gamma$, $\gamma=1,\ldots,M_{\rm e}\}$
are a set of auxiliary variables known as spin-rapidities. 
The scattering phase shifts are given by $\theta(x) = 2\tan^{-1}(x/c)$,
with $c=2\lambda/(1-3\lambda^2/4)$ playing the role of an effective 
coupling constant.
We have also defined $f_{c,s} = (f_\uparrow \pm f_\downarrow)/2$. 
The first BAE simply gives the quantum numbers of free, chiral
electrons, written in the Bethe ansatz basis.
The second BAE describes the 
charge degrees of freedom in the even channel (holons), while 
the third  describes
the spin degrees of freedom in the even channel (spinons). 

Consider first the case where only a magnetic flux threads the ring,
with a spin-independent integrable AB phase 
$f_{\uparrow} = f_{\downarrow}=f$. To be explicit, we choose $N=4n+2$ with
$M_{\rm e/o} = (N_{\rm e/o}\pm 1)/2$, where $N_{\rm e}=N_{\rm o}$. 
For this case, the groundstate is
characterized by integer-spaced quantum numbers $\{n_l, J_\delta,
n_r, I_\gamma\}$ in the symmetric
ranges $-(N_{\rm o}-1)/2 \leq n_l \leq (N_{\rm o}-1)/2$,
$-(M_{\rm o} - 1)/2 \leq J_\delta \leq (M_{\rm o} - 1)/2$,
$-(N_{\rm e}-1)/2 \leq n_r \leq (N_{\rm e}-1)/2$, and
$-(M_{\rm e} - 1)/2 \leq I_\gamma \leq (M_{\rm e} - 1)/2$.
Once a set of
spin rapidities $\Lambda_\gamma$ is given,
the momenta $k_{n_r}$ are determined, and the excess numbers on the
right and left branches can be derived.
With $f_c=f$
and $f_s=0$, it becomes immediately clear from the BAE that the 
total scattering phase 
shift from the dot is {\em independent} of $f$,
and we thus obtain 
that $\Delta N_r(f) = - \Delta N_l(f) = f/2$.
For an even number of
electrons, the sum in (\ref{PersistentCurrent})
can be written as $2(\Delta N_{r}(f_c,f_s) - \Delta
N_{l}(f_c,f_s))$, 
and we obtain 
$I(\Phi,0) = -(4ev_{\rm F}/L)(\Phi/\Phi_0)$ at the integrable points 
$\Phi/\Phi_0 = f/2$.
This is the same PC as that for an ideal ring of free 
electrons.
The analogous analysis for $N=4n, N=4n+1$ and $N=4n+3$ reveals that for 
all cases
{\em the AB PC is unaffected by the Kondo resonance
at the integrable points} $\Phi/\Phi_0 = f/2$. 

It should be pointed out that the PC we have calculated is the current
carried by the state which evolves adiabatically from the ground state
at $\Phi=0$ as $\Phi$ is turned on.
Generically, there may be one or
more level crossings \cite{casajm} between $f=0$ and $f=1$,  
so that this state may not be the absolute ground state at e.g.\ $f=1$.
Notwithstanding this proviso, it is clear from the following argument 
that the above result for $I(\Phi,0)$ 
must hold for small values of $\Phi/\Phi_0$:
By symmetry, the PC is an odd function of $\Phi$ and is
analytic, except at values of $\Phi$ corresponding to level
crossings. Considering again the case $N=4n+2$, there is no level
crossing at $\Phi=0$.  The leading mesoscopic behavior of the PC is
then $ I(\Phi,0) = -D_c \Phi/L + {\cal O}(\Phi^3/L^3)$, where $D_c$ is
the charge stiffness.  This holds on general grounds, independent of
whether the model is integrable or not.  
To determine the charge stiffness, however, we only need to consider the
state which evolves adiabatically \cite{haldane}
from the ground state at $f=0$,
specified above, as $\Phi$ is increased.  Thus the above result for 
$I(\Phi,0)$ is seen to hold for small values of $\Phi/\Phi_0$ as well.

Our result for the AB PC suggests that spin-charge separation
holds generically at the mesoscopic scale of this system.
The phase shift acquired
by scattering off the dot in the even parity channel affects only the spin 
content of an incoming electron \cite{baKondo},
but not the charge which contributes to the PC. This may explain
why our result for a ring is different from that obtained for a sidebranch
dot connected to an open quantum wire  \cite{Kang2}: there it was found
that the $\pi/2$ phase shift due to the Kondo resonance suppresses
the transmission completely. However, in an open system where
single electrons can be added and subtracted (via a source and
drain), 
the holons and spinons - which are the relevant low-energy degrees
of freedom in
the correlated BA basis - 
have to be recombined in such a way as to mimic single-electron
excitations,
and this makes the transmission susceptible to spinon scattering
\cite{FrojdhJohannesson,cas}. 
No such constraint is expected to apply to a correlated {\em closed}
system,
and the AB PC - carried by the holons - can flow unperturbed by
scattering 
events in the spin sector.  

What then happens to the
PC due to a  
{\em spin-dependent} AC flux? Although dynamical spin-charge separation still
holds at the integrable points---as is evident from the one-dimensional
Kondo Hamiltonian in (\ref{kondo})---the 
spin-dependent phase induced by the AC flux may now carry information 
from the
spin sector into the charge sector, hence affecting the PC.
The analysis of the BAE
in the case of general spin--dependent
fluxes (however, still satisfying $\phi_\alpha=f_\alpha\pi$ with $f_\alpha$ an
integer) is more involved. 
However, in the strong-coupling limit $c \rightarrow \infty$ 
and with a pure 
AC flux (i.e.\ with $f_{\uparrow} = - f_{\downarrow} = f$)
we can still obtain an analytic solution. For this case,
$\theta(2\Lambda_\gamma - 2) \rightarrow \theta(2\Lambda_\gamma)$ 
in the BAE, and one may algebraically solve for the
excess numbers.
Again specializing to the case 
$2N_{\rm e}=2N_{\rm o}=N=4n+2$ with   
$M_{\rm e/o} = (N_{\rm e/o}\pm 1)/2$, we obtain 
\begin{equation}
I(0, \tau) =
 2\frac{e v_{\rm F}}{L}\cdot\frac{\tau}{\tau_0} \left(\frac{1}{N_{\rm
o}}\right), \ \ \ \ c \rightarrow \infty. \label{ACcurrent}
\end{equation}
Since $v_F \sim N_{\rm o}$, the current $I(0, \tau)$
is {\em independent of the number of electrons in the ring}, and only depends 
on the charge density $\tau$ and the ring's circumference $L$.
Eq.\ (\ref{ACcurrent}) is to be compared to the case of an
ideal ring with an equal number of
up- and down-spin electrons, for which case there is {\em no} 
AC PC \cite{BA}.
This is easily verified in the present formalism by simply removing the 
quantum dot scattering
term $\theta(2\Lambda_\gamma)$ from the BAE.
Solving for the PC one obtains 
$I(0, \tau) = -(2ev_{\rm F}/L)(\tau/\tau_0)(N_{\rm e}^{-1} - N_{\rm o}^{-1})
= 0$.

The result in (\ref{ACcurrent}) strikingly exhibits the nondynamical 
nature of the AC effect:
In the strong-coupling limit the dot effectively screens the 
spin content of an electron, and via the AC boundary condition this
causes the charge of an electron in the even parity, 
right-moving sector to get trapped. 
It is crucial to realize that
{\em this happens in the absence of any interaction
term in the Hamiltonian containing the charge density}.
As a result, the numbers of current carrying
states on the right and left branches become unbalanced, 
leading to a net AC persistent
current.

In conclusion, we have shown how a one-dimensional mesoscopic ring 
of free electrons
coupled by a tunnel junction to a quantum dot at a Kondo resonance  
can be described
by an exactly solvable model for certain privileged values of the
Aharonov-Bohm-Casher fluxes piercing the ring. 
This allows for an exact nonperturbative analysis of the relevant
physics, in particular how the PC is influenced by a Kondo resonance.

We thank N.\ Andrei, R.\ Egger, H.\ Grabert and A.\ Zvyagin for discussions.
This work was supported in part by the Finnish
Academy and the Deutscher Akademischer Austauschdienst.
HPE acknowledges support from the Australian Research Council,
HJ from the Swedish Natural Science 
Research Council, and CAS from NSF grant DMR0072703.


\vspace*{-3mm}

\begin{references}

\vspace*{-13mm}

\bibitem{Hewson} See e.g.\ A.\ C.\ Hewson,
                 {\em The Kondo Problem to Heavy Fermions}
                 (Cambridge University Press, Cambridge, 1993). 

\bibitem{Goldhaber} D.\ Goldhaber-Gordon {\em et al.}, 
                    Nature {\bf 391}, 156 (1998);
                    S.\ M.\ Cronenwett {\em et al.}, 
                    Science {\bf 281}, 540 (1998);
                    F.\ Simmel {\em et al.},
                    Phys.\ Rev.\ Lett.\ {\bf 83}, 804 (1999);
W. G. van der Wiel {\em et al.}, Science {\bf 289}, 2105 (2000).

\bibitem{Glazman} L.\ I.\ Glazman and M.\ E.\ Raikh, 
                  JETP Lett.\ {\bf 47}, 452 (1988); 
                  T.\ K.\ Ng and P.\ A.\ Lee, 
                  Phys.\ Rev.\ Lett.\ {\bf 61}, 1768 (1988).

\bibitem{Meir} Y. Meir, N. S. Wingreen, and P. A. Lee, 
Phys. Rev. Lett. {\bf 70}, 2601 (1993); N. S. Wingreen and Y. Meir,
Phys. Rev. B {\bf 49}, 11040 (1994).


\bibitem{by} N.\ Byers and C.\ N.\ Yang,
             Phys.\ Rev.\ Lett.\ {\bf 7}, 46 (1961).

\bibitem{BIL} M.\ B\"{u}ttiker {\em et al.},
              Phys.\ Lett.\ {\bf A96}, 365 (1983).

\bibitem{ZK} For a review, 
             see e.g.\ A.\ A.\ Zvyagin and I.\ V.\ Krive, 
             Low Temp.\ Phys.\ {\bf 21}, 533 (1995).

\bibitem{AB} Y.\ Aharonov and D.\ Bohm, Phys.\ Rev.\ {\bf 115}, 485 (1959).

\bibitem{AC} Y.\ Aharonov and A.\ Casher, 
             Phys.\ Rev.\ Lett.\ {\bf 53}, 319 (1984).

\bibitem{Yacoby} A.\ Yacoby, {\em et al.}, 
                 Phys.\ Rev.\ Lett.\ {\bf 74}, 4047 (1995).

\bibitem{mbcs} M.\ B\"uttiker and C.\ A.\ Stafford, 
               Phys.\ Rev.\ Lett.\ {\bf 76}, 495 (1996).

\bibitem{Ferrari} V.\ Ferrari {\em et al.},
                  Phys. Rev. Lett. {\bf 82}, 5088 (1999).

\bibitem{Kang} K.\ Kang and S.-C.\ Shin, Phys. Rev. Lett. {\bf 85}, 5619 (2000).

\bibitem{Deo} P.\ S.\ Deo and A.\ M.\ Jayannavar, cond-mat/0006035.

\bibitem{ZB} A.\ A.\ Zvyagin and T.\ V.\ Bandos,
             Low Temp. Phys. {\bf 20}, 222 (1994).

\bibitem{Cheung} H.-F. \ Cheung {\em et al.}, Phys. Rev. B {\bf 37}, 6050 (1988).

\bibitem{Loss}  D.\ Loss, Phys.\ Rev.\ Lett.\ {\bf 69}, 343 (1992); A. \ O. Gogolin and
                N. \ V. Prokof'ev, Phys.\ Rev.\ {\bf B} 50, 4921 (1994).

\bibitem{footnote} When $U$ decreases
               below $W$, the band width of the electrons in the ring, it just trades place with
               $W$ in the scaling expresion for the Kondo temperature, but otherwise
               does not affect the relevant physics 
(see Ref. \cite{Meir}).

\bibitem{Thimm} W. \ B. \ Thimm, J. Kroha, and J. von Delft, Phys. Rev. Lett. {\bf 82}, 2143 (1999).

\bibitem{baKondo} For a recent review, 
                  see e.g.\ N.\ Andrei
                  in {\em Series on Modern Condensed Matter Physics -
                  Vol. 6}, 458
                  (World Scientific, Singapore, 1992), Eds. S.\ Lundquist,
                  {\em et al.}

\bibitem{we} F.\ Woynarovich and H.-P.\ Eckle, 
             J.\ Phys.\ A: Math.\ Gen.\ {\bf 20}, L443 (1987); 
             H.-P.\ Eckle, 
             in {\em Strongly interacting fermions and high--$T_c$
                 superconductivity},
             B.\ Dou\c{c}ot and J.\ Zinn--Justin, eds.,
             Proc.\ Les Houches 1991, Session LVI,
             (North--Holland, Amsterdam, 1995).

\bibitem{casajm} N.\ Yu and M.\ Fowler,
                 Phys.\ Rev.\ B {\bf 45}, 11795 (1992);
                 C.\ A.\ Stafford and A.\ J.\ Millis,
                 Phys.\ Rev.\ B {\bf 48}, 1409 (1993). 

\bibitem{haldane} F.\ D.\ M.\ Haldane, Phys. Lett. {\bf 81A}, 153 (1981).

\bibitem{Kang2} K.\ Kang, {\em et al.}, cond-mat/0009235.

\bibitem{FrojdhJohannesson} P.\ Fr\"ojdh and H.\ Johannesson,
                            Phys.\ Rev.\ Lett.\ {\bf 75}, 300 (1995); 
                            Phys.\ Rev.\ B {\bf 53}, 3211 (1996). 

\bibitem{cas} C.\ A.\ Stafford, Phys.\ Rev.\ B {\bf 48}, 8430 (1993).

\bibitem{BA} A.\ V.\ Balatsky and B.\ L.\ Altshuler, 
             Phys. Rev. Lett. {\bf 70}, 1678 (1993).

\end{references}
\end{document}